\documentclass[prb, showpacs, superscriptaddress]{revtex4}
\usepackage{amssymb}
\usepackage{wasysym}
\usepackage{graphicx}

\begin{document}

%-----------------------------------------------------------------------

\title{Extrinsic faulting in $3C$ close packed crystal structures: Computational mechanics analysis}

%------------------------------------------------------------------------

\author{Ernesto \surname{Estevez-Rams}}
\affiliation{Facultad de F\'isica-IMRE, Universidad de la Habana, San Lazaro y L. CP 10400. C. Habana. Cuba}
\email{estevez@fisica.uh.cu}
\author{Raimundo \surname{Lora-Serrano}}
\affiliation{Universidade Federal de Uberlandia, AV. Joao Naves de Avila, 2121- Campus Santa Monica, CEP 38408-144, Minas Gerais, Brazil}
\author{Arbelio \surname{Penton-Madrigal}}
\affiliation{Facultad de F\'isica-IMRE, Universidad de la Habana, San Lazaro y L. CP 10400. C. Habana. Cuba}
\author{Massimo \surname{Nespolo}}
\affiliation{Universit\'e de Lorraine, CRM2, UMR 7036, Vandoeuvre-l\'es-Nancy, F-54506 France}

\date{\today}

%---------------------------------------------------------------------------
\begin{abstract}
Extrinsic faulting has been discussed previously within the so called difference method and random walk calculation. In this contribution is revisited under the framework of computational mechanics, which allows to derive expressions for the statistical complexity, entropy density and excess entropy as a function of faulting probability. The approach allows to compare the disordering process of extrinsic fault with other faulting types. The $\epsilon$-machine description of the faulting mechanics is presented. Several useful analytical expressions such as probability of consecutive symbols in the H\"agg coding is presented, as well as hexagonality. The analytical expression for the pairwise correlation function of the layers is derived and compared to results previously reported. The effect of faulting on the interference function is discussed as related to the diffraction pattern. 
\end{abstract}


\maketitle

%----------------------------------------------------------------------------------------------------------------
\section{Introduction}
%-----------------------------------------------------------------------------------------------------------------

Close packed structures (CPS) are OD (Order-Disorder) structures built by stacking hexagonal layers in the direction perpendicular to the layer \cite{durovic97}. The stacking ambiguity raising from the two possible positions of a layer with respect to the previous one leads to a theoretically infinite number of possible polytypes if no constraint is made on the periodicity. However, by far the commonest ones are the cubic close packed or $3C$  (Bravais lattice of type $cF$) and the hexagonal close packed or $2H$ (Bravais lattice of type $hP$). These are MDO (Maximum Degree of Order) polytypes, meaning that they structure contains the minimal number of layer triples, quadruples etc. (one, in both cases). CPS usually exhibit some kind of planar disorder, or stacking faults, that viewed as a disruption in the otherwise periodic arrangement of layers, can be analyzed as non interacting defects. This is the basis of the so called random faulting model (RFM), that has been the most widely used model of faulting in layer structures, dating back to the early times in diffraction analysis  \cite{wilson,wagner,warren}. The idea  of the RFM is to consider certain types of faulting, such as intrinsic (removal of a layer from the sequence), extrinsic (addition of a layer in the sequence) and twinning (change of orientation in the sequence), assigning to each a fixed probability of occurrence, independent from the density of faulting and neglecting any spatial interaction between the faults present in the material. This simplifying assumption means that RFM, if suitable, should be at very low density of faulting, in which case it is justified, during the derivation of the correspondent expression for the displacement correlation function between layers (also known as the pairwise correlation) and subsequently the diffraction equation, to drop all terms above linear in the faulting probabilities. Analytical expressions are then found: see such mathematical development in the classical works of \cite{warren} for deformation and twin faulting in $FCC$ and $HCP$ structures. 

Besides the assumption of low density of defects, the RFM also assumes that faults, when occur, go through the whole coherently diffracting domain, avoiding the need to account for the appearance of partial dislocations. Further, faulting are considered to happen along the stacking direction but not along any direction that is crystallographically equivalent in the non-faulted polytype. For example, in the case of the unfaulted $3C$ polytype, the four directions $\langle 111 \rangle$ are equivalent, as in any cubic crystal, but this is no longer the case if faulting occurs in one of the four. The reader can refer to \cite{estevez07} for further historical account of the subject. 

In recent years there have been attempts to extend the mathematical applicability of the model, without modifying its fundamental assumptions. First, \cite{velterop} have observed that, even for low density of faulting, the assumption of only one faulting direction, is an unrealistically simplifying assessment of the diffraction behavior, which can lead to misleading conclusions. Another issue is the need to accommodate larger, more realistic, faulting densities within the model. Even if the physical assumption of non interacting faulting is too heavy for larger probability of faulting, it is still interesting to pursue such an extension for several reasons. RFM can be used as a reference model for other approaches. The fact that only one parameter for each faulting type is needed, makes it very attractive in practical analysis of materials. Additionally, RFM can be used as a suitable starting model in computer simulations of faulting. In this case, the need of a good starting proposal is essential in the convergence and convergence speed of numerical calculations.  

In the last years, independently,  Varn \textit{et al.} \cite{varn01,varn01a,varn02,varn04}, and \cite{estevez08}, have attempted to rewrite the RFM in a modern framework, using a Hidden Markov Model (HMM) description of the faulting dynamics. The more ambitious idea is to go beyond the faulting model and try to understand the disordering process in layer structures, as a dynamical process of a system capable of performing (physical) computation and, in this sense, able to store and process information \cite{crutchfield92}. The attractiveness of the proposal is that such approach can harvest from a powerful set of tools, developed within the study of complexity, grounded in information theory concepts such as Shannon entropy and mutual information. This framework  is known as computational mechanics and have been used in a wide range of subjects \cite{crutchfield12}.

A first attempt to use the HMM description of random faulting was done by \cite{estevez08} for intrinsic and twinning faults. Their analysis allowed to calculate the displacement correlation function and the diffraction equation for the whole range of faulting probabilities. They also derived useful expression concerning the hexagonality of the stacking event, the average size of cubic and hexagonal neighborhood blocks, the correlation length, all as function of the faulting probabilities.  While correct, this approach is ad hoc in nature, only applicable to the problem considered by them, namely using as starting structure the $3C$ layer ordering, and working through the appropriate equations. 

A more recent breakthrough came from the work of \cite{riechers15}, who proved that calculation of the pairwise correlation function could be systematized in an elegant way, allowing its applicability to a wide number of situations such as those found in close packed arrangements. The idea is to find the description of the stacking arrangement as a HMM and, from there, build the transition matrix, find the stationary probabilities of the HMM states and the pairwise correlation function [See equation (13) in \cite{riechers15} or in this contribution further on]. In their contribution, they also discussed a number of examples that showed how the formalism can reproduce previous results, such as those reported by \cite{estevez08} and also be applied to other cases. 

The result of \cite{riechers15} opens the possibility to study, in a systematic way, the RFM for different types of faulting and their combinations, something which proved to be at least cumbersome and, in certain instances, intractable by previous tools. This is what we intend to do in this contribution for extrinsic faults.  Extrinsic faulting has been dealt before \cite{johnson63,warren63,lele67,holloway69,holloway69a,howard77,takahashi78,howard79}.

The main goal of the manuscript is to report several analytical expressions for disorder of extrinsic faulted CPS. These expressions relates disorder magnitudes such as those derived within mechanical computation with the extrinsic faulting probability which in turn allows comparison with similar expression derived for twin and deformation fault already reported \cite{estevez08}. Also, closed analytical expression for the probability of finding different stacking sequences in the faulted structure is reported and from there an expression is derived for the hexagonality and the average length of perfectly coherent $FCC$ sequence within the CPS. The analytical expression of the pair correlation function as a function of faulting probability is derived and its decaying and oscillation behavior are discussed. Finally, the expression for the interference function is reported and peak shift and asymmetry as a result of the extrinsic faulting is commented.

First the main concepts used and the notation are introduced.

%----------------------------------------------------------------------------------------------------------------
\section{Order and disorder in close packed stacking arrangements and the pairwise correlation function}
%-----------------------------------------------------------------------------------------------------------------

In the OD structures built from hexagonal layers, the layers can be found only in three positions perpendicular to the stacking direction, which are usually labeled $A$, $B$, and $C$ \cite{durovic97,pandey2}. Close packed is the constraint that two layers  which bear the same letter, and are thus exactly overlapped in the projection along the stacking direction can not occur consecutively.  According to this description, the ideal $FCC$ structure is described by $ABCABCAB\ldots$  sequences \cite{verma}, while the ideal $HCP$ structure, has a stacking order described by $ABABABA\ldots$ and the double hexagonal close packed ($DHCP$) in turn is described by the stacking $ABCBABCB\ldots$. 

An equivalent, and less redundant coding is the H\"agg code \cite{hagg43}, where pair of consecutive layers is given a plus (or 1) symbol if they form a ''forward'' sequence $AB$, $BC$ or $CA$, and a minus (or 0) sign otherwise\footnote{An alternative notation by Nabarro-Frank \cite{frank51} uses $\bigtriangledown$ and $\bigtriangleup$ for $+$ (or 1) and $-$ (or 0) respectively.}. There is a one-to-one relation between both coding \cite{estevez05a}. It is also important to introduce a three layer hexagonal environment as one where a layer $X$ has the two adjacent layers in the same position (e.g. $ABA$, $ACA$, $BAB$, $BCB$, $CAC$, $CBC$); if a layer environment is not hexagonal then it is cubic. A hexagonal environment is denoted by a letter $h$ and a cubic environment by a letter $k$, this is the basis of the Jagodzinski coding of the stacking arrangement, as before, there is a one-to-one correspondence between the $ABC$ coding and the Jagodzinski coding \cite{estevez08a}. Hexagonality then refers to the fraction of hexagonal environments in the stacking sequence. Also, it can be easily checked that, when in the H\"agg code the pair of characters $10$ or $01$ is found, a hexagonal environment is found.

Faulting is generically meant as a disruption of the ideal periodic ordering of a stacking arrangement and therefore constitute a defect in the structure. In close packed structures, the most simple type of faults that are usually considered are (1) deformation faults, which are jogs  in the otherwise perfect periodic sequence, (2) extrinsic or double-deformation fault, which is the insertion of an extraneous layer in the perfect sequence and; (3) twin faults, which cause reversions in the stacking ordering. In what follows, the probability for the occurrence of a deformation fault will be denoted by $\alpha$, of an extrinsic fault by $\gamma$, while the probability for the occurrence of a twin faulting will be denoted by $\beta$.

The pair correlation function between layers, known as the pairwise correlation function $Q_{\xi}(\Delta)$, is the key to calculate the effect of the stacking arrangement in the diffraction intensity \cite{estevez01,estevez03a}. Consider a stacking direction and sense, $Q_{\xi}(\Delta)$, where $\xi=\{c,a,s\}$ is the probability of finding two layers, $\Delta$ layers apart, and displaced the first with respect to the second one as (1) $\xi=c$: $A-B$ or $B-C$, or $C-A$; (2) $\xi=a$: $B-A$ or $C-B$, or $A-C$ and (3) $\xi=s$: $A-A$ or $B-B$, or $C-C$\footnote{The notation is that used by Varn et al. \cite{varn01,varn02}, where $c$ stands for ''cyclic'', $a$ stands for ''anti-cyclic and $s$ for ''same''}. It should be noted that $Q_{s}(1)=0$ due to the close packed constraint and $Q_{s}(0)=1$ by construction. 

\begin{figure}
\caption{The HMM in the H\"agg notation for the perfect FCC structure $\ldots 11111 \ldots$; the HCP structure $\ldots 1010101 \ldots$; and DHCP structure $\ldots 1100110011 \ldots$. Labels at the transition arcs are of the form $s|p$ where $s$ is the emitted symbol and $p$ the probability of making the transition from one state to the other.}
\scalebox{1.2}{\includegraphics{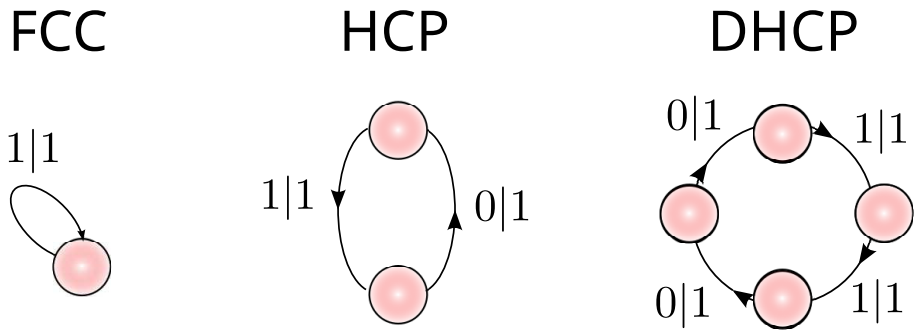}}\label{fig:perfseq}
\end{figure}

It is possible, for any of the described codings ABC, H\"agg and Jagodzinski, to construct a Hidden Markov Model (HMM) describing a broad range of both ordered and disordered stacking process. A HMM description comprises a finite, or at least enumerable, set of states $\mathcal{S}$ and the associated initial set of probability $\pi_0$ of being in each state; the set of transition matrices $\mathbf{T}$, and a set of symbols drawn from a finite alphabet $\mathcal{A}$. Each transition matrix  $\mathcal{T}^{[\upsilon]}$ is a square matrix with number of rows equal to the number of states, where each entry $t^{(\upsilon)}_{ij}$ represents the probability of jumping from state $i$ to state $j$, while emitting the symbol $\upsilon\in\mathcal{A}$. The HMM transition matrix $\mathcal{T}$ is defined as the sum of the $\mathcal{T}^{(\upsilon)}$ over all symbols $\upsilon$ in the alphabet. Figure \ref{fig:perfseq} shows the HMM for the $FCC$, $HCP$ and $DHCP$ stacking structures. For further details the reader is referred to previous papers on the subject \cite{varn01,varn01a,varn02,varn04,estevez08}. 

When seen through a HMM description, stacking arrangements are cast as an information processing system that sequentially outputs symbols as it makes transition between states. The system output is then an infinite string of characters $\Upsilon=\ldots \upsilon_{-2} \upsilon_{-1} \upsilon_{0} \upsilon_{1} \upsilon_{2}\ldots$ each character $\upsilon_i\in\mathcal {A}$. For the purposes of analysis it is common to, at a given point, divide the output string in two halves, the left halves $\overleftarrow{\Upsilon}=\ldots \upsilon_{-2} \upsilon_{-1}$ is known as the past, while the right halve $\overrightarrow{\Upsilon}=\upsilon_{0} \upsilon_{1} \upsilon_{2}\ldots$ is called the future \footnote{The terms of past and future are taken from the analysis usually carried out in dynamical systems and is kept even when the considered variable is not time, as in the case of stacking order where the pertinent variable is layer position in the stacking. In any case, stacking and faulting are usually cast as a sequential process \cite{warren63}, the HMM analysis just makes this explicit. One could understand the meaning of past and future in this sense.} \cite{varn01,varn01a,varn02,varn04}. There can be many HMM describing the same process, the minimal HMM describing the system dynamics is considered to be optimal in the sense of using fewer resources while providing the best predictive power and will be the one relevant in this contribution, such model is called an $\epsilon$-machine \cite{crutchfield92,crutchfield12}.  The $\epsilon$-machine has, among others, the important property of unifilarity which means that, from a given state, the emitted symbols determines unambiguously the transition to another state.

Let us denote, following the common use of brac and kets in physics, by $\langle \pi |$ the vector of state probabilities and by $|1\rangle$ a vector of $1s$. If the HMM description is known, then the probability of any finite sequence $\upsilon^N=\upsilon_i \upsilon_{i+1} \upsilon_{i+2}\ldots \upsilon_{i+N-1}$ will be given by
\begin{equation}
 P(\upsilon^N)=\langle \pi | \mathcal{T}^{[\upsilon_{i}]} \mathcal{T}^{[\upsilon_{i+1}]}\ldots\mathcal{T}^{[\upsilon_{i+N-1}]}|1\rangle.\label{eq:psn}
\end{equation}
Where in this case $\langle x | A | y\rangle$ is a real number resulting from the scalar product between vectors and matrices.

Several information theory magnitudes can be defined once the minimal HMM  description of the stacking process is known. Shannon defined information entropy $H(X)$ for an event set $X$ with discrete probabilities distribution $p(X)$ as \cite{arndt}
\begin{equation} 
 H(X)=-\sum_i p(X) \log p(X),
\end{equation}
where the sum is taken over all the probability distribution and here and in what follows the logarithm is taken base two which makes the units of the entropy to be bits.

For the $\epsilon$-machine, the statistical complexity $C_\mu$ is defined as the Shannon entropy over the HMM states,
\begin{equation}
 C_\mu=H(\mathcal{S})=-\sum_i p_i \log p_i,\label{eq:sc} 
\end{equation}
where $p_i$ is the stationary probability of the $i$th-state in the minimal HMM description and the sum is over all states probabilities. $C_\mu$ measures the amount of information the system stores. 

Excess entropy $E$ is also used to characterize the system information processing capabilities and is used as a measure of predictability, defined as the mutual information between the left half and the right half in the system output,
\begin{equation}
 E=H(\overleftarrow{\Upsilon})+H(\overrightarrow{\Upsilon})-H(\Upsilon). 
\end{equation}

Entropy density $h_\mu$ \cite{arndt} is defined as 
\begin{equation}
h_\mu=\lim_{N\rightarrow\infty}\frac{H(\Upsilon^N)}{N},
\end{equation}
when such limit exist,  with $\Upsilon^N$denotes all substrings of $\Upsilon$ of length $N$. $h_\mu$ is used to answer how random the process is \cite{feldman03}.

Finally, \cite{riechers15} described a procedure for computing the pairwise correlation function from the transition matrices, that can be summarized as follows:
\begin{enumerate}
 \item The HMM of the stacking process in the ABC notation is given together with $\{\mathcal{A}, \mathcal{S}, \pi_0, \mathbf{T}\}$. If this description is given in the H\"agg coding then the expansion to the ABC coding must be performed \cite{riechers15}.
 \item The stationary probabilities $\pi$ over the HMM states is calculated as the normalized left eigenvector of the transition matrix $\mathcal{T}$ with eigenvalue unity.
 \begin{equation}
 \langle\pi|=\langle\pi|\mathcal{T},\label{eq:eigen}
\end{equation}
 \item The pairwise correlation function follows from the definition and the use of equation (\ref{eq:psn}): 
 \begin{equation}
  \displaystyle Q_{\xi}(\Delta)=\sum_{x_0\in\mathcal{A}}\langle \pi | \mathcal{T}^{[x_0]}\mathcal{T}^{\Delta-1}\mathcal{T}^{[\hat{\xi}(x_0)]}|\mathbf{1}\rangle.\label{eq:Q}
 \end{equation}
Where $\hat{\xi}=\{\hat{c}, \hat{a}, \hat{s}\}$ is a family of permutation functions given by
\begin{equation}
\begin{array}{lll}
\hat{c}(A)=B & \hat{c}(B)=C & \hat{c}(C)=A\\
\hat{a}(A)=C & \hat{a}(B)=A & \hat{a}(C)=B\\
\hat{s}(A)=A & \hat{s}(B)=B & \hat{s}(C)=C
\end{array}
\end{equation}
 and $\mathbf{1}$ represents a vector of $1$'s (See also equations (20) and (24) in \cite{riechers15} for alternatives expression for equation (\ref{eq:Q})).
\end{enumerate}

%-------------------------------------------------------------------------------------------------------------------
\section{Extrinsic fault in the face centered cubic stacking order}
%-------------------------------------------------------------------------------------------------------------------

\begin{figure}
\caption{Extrinsic faulting in the sequence $3C$ along the cubic [111] direction. An ideal periodic sequence is shown for comparison.}
\scalebox{0.8}{\includegraphics{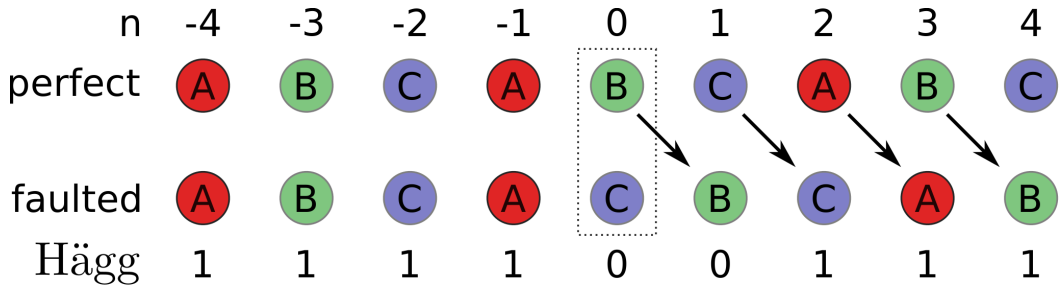}}\label{fig:ef}
\end{figure}

An extrinsic fault (In the case of the FCC structure also known as double deformation fault) in a $3C$ stacking is depicted in Fig. \ref{fig:ef} along a perfect sequence for comparison. It can be seen that in the H\"agg code, the extrinsic fault is equivalent to the flip (bitwise negation) of two consecutive characters. The probability of occurrence of such faulting will be denoted by $\gamma$. It will be assumed that $\gamma$ can take any value between 0 and 1. Building from the effect of the extrinsic fault over the H\"agg code, the HMM of the faulting process is shown in Fig. \ref{fig:fsaext}, where it is assumed that the ideal $3C$ structure goes in the $A\rightarrow B \rightarrow C$ sequence. The $p$ state represents the non faulted condition, as long as the system stays in that state, the output symbol $\upsilon=1$ will correspond to the perfect $3C$ structure. If faulting occurs, a $0$ is emitted and the system goes to the $e$ state, where a second $0$ is printed with certainty while returning to the $p$ state\footnote{The described dynamics implicitly assumes that an inserted layer can not follow another inserted layer. The later case has been approached by \cite{howard77}.}. The HMM of figure \ref{fig:fsaext} represents a biased even process (see Appendix A of \cite{crutchfield13} and Example D in \cite{varn13}). 

It should be observed that any sequence with an odd number of $0$ can not be the result of such HMM. Such sequence will be called forbidden, moreover, the forbidden sequences are called irreducible as they do not contain a proper subsequence which itself is forbidden. The number of irreducible forbidden sequence in the even process is infinite, in such case, the process is called a sofic system \cite{feldman03}. The fact that any sequence from the HMM of the even process contains an even number of $0's$ has important consequences as will be discussed further down. 

The corresponding transition matrix will be given by
\begin{equation}
 \begin{array}{ll}
  \mathcal{T}^{[1]}=\left ( \begin{array}{ll}\overline{\gamma}&0\\0&0\end{array} \right ) & \mathcal{T}^{[0]}=\left ( \begin{array}{ll}0&\gamma\\1&0\end{array} \right ) \\\\
  \mathcal{T}=\left ( \begin{array}{ll}\overline{\gamma}&\gamma\\1&0\end{array} \right ), &
 \end{array}
\end{equation}
where $\overline{\gamma}$ stands for $1-\gamma$. The stationary probabilities over the recurrent states $p$ and $e$ can be calculated following equation (\ref{eq:eigen}) which results in 
\begin{equation}
 \displaystyle \langle\pi|=\left \{\frac{1}{1+\gamma},\frac{\gamma}{1+\gamma}\right \},\label{eq:pi}
\end{equation}
the first value corresponds to the $p$ state.

Hexagonality in terms of computational mechanics has been analyzed in a more general context previously \cite{varn07}. Hexagonality can be calculated from the probability of occurrence of $01$ or $10$ in the H\"agg code of the sequence. Both probabilities are equal and, from equation (\ref{eq:psn}),  given by
\begin{equation}
 P(01)=\langle \pi | \mathcal{T}^{[0]} \mathcal{T}^{[1]}|1\rangle=\gamma \frac{1-\gamma}{1+\gamma},
\end{equation}
from which the hexagonality is given by $2P(01)$. Hexagonality has a maximum value of $2(3-2\sqrt{2})\approx 0.343$ at $\gamma=\sqrt{2}-1\approx 0.414$ (Fig. \ref{fig:hexhe}a).

The statistical complexity can be derived from equation (\ref{eq:sc}) using equation (\ref{eq:pi}) and is given by
\begin{equation}
 \displaystyle C_{\mu}=\frac{1}{1+\gamma}\left ( \log (1+\gamma)-\gamma \log \frac{\gamma}{1+\gamma}\right ).
 \end{equation}
logarithm is taken usually in base two and then the units of $C_\mu$ is bits.  For an $\epsilon$-machine the entropy density is given by \cite{crutchfield13}
 \begin{equation}
  \displaystyle h_\mu=-\sum_{k\in \mathcal{S}}P(k)\sum_{x \in \mathcal{A}} P(x|k)\log P(x|k),
 \end{equation}
 where $P(a|b)$ means the probability of $a$ conditioned on $b$. The units of the entropy density is bits/site. The expression for the entropy density will not be derived explicitly and the reader is referred to \cite{crutchfield13}, the resulting expression for the entropy density is
 \begin{equation}
  \displaystyle h_{\mu}=-\frac{1}{1+\gamma}\left [ \gamma\log \gamma+(1-\gamma)\log (1-\gamma)\right ].
 \end{equation}
 
The calculation of the excess entropy is more involved and explained in detail in the Appendix. The results is
 \begin{equation}
 \displaystyle E=\frac{1}{1+\gamma}\left ( \log (1+\gamma)-\gamma \log \frac{\gamma}{1+\gamma}\right ),
 \end{equation}
which is identical to the statistical complexity. 

Figure \ref{fig:hexhe}b shows the behavior of the excess entropy as a function of $\gamma$. Observe that at $\gamma=1$, the excess entropy has a discontinuity, as $E$ drops to zero when the finite state automata description has a topological change to a certain process with only one state and emitting always a $0$ symbol. This discontinuity is not seen by the entropy density (Fig. \ref{fig:hexhe}a) that has a maximum at $\gamma=1/2(3-\sqrt{5})\approx 0.382$ with value $h_\mu=0.6942$ bits/site and then smoothly drops to zero as $\gamma$ approaches $1$.

The probability of a chain of 0's of length $n$ is given by
\begin{equation}
 P(0^n)=\left \{ \begin{array}{ll}\gamma^l\left(1-\frac{2 \sqrt{\gamma}}{1+\gamma}\right)& n=2 l\\0 & n=2l+1\end{array}\right. .\label{eq:0}
\end{equation}
For chains of 1's
\begin{equation}
 P(1^n)=\frac{(1-\gamma)^n}{1+\gamma}.\label{eq:1}
\end{equation}
From equation (\ref{eq:0}) and (\ref{eq:1}) the average length of blocks of 0's and 1's can be calculated
\begin{equation}
 \begin{array}{l}
  \langle L_0 \rangle=\sum_{n=1}^{\infty}n P(0^n)=\frac{4 \gamma}{(1-\gamma)^2}\\
  \\
  \langle L_1\rangle=\sum_{n=1}^{\infty}n P(1^n)=\frac{1}{\gamma^2}\frac{1-\gamma}{1+\gamma}.
 \end{array}
\end{equation}
$\langle L_0\rangle=\langle L_1\rangle$ at $\gamma=0.3623$.

In Fig. \ref{fig:hexvse} hexagonality as a function of excess entropy and entropy density are shown. The higher the entropy density is, the higher the hexagonality,
which comes as no surprise as hexagonal neighborhoods are result of faulting events, which in turn implies larger disorder. It can be seen though, that hexagonality is not a function of entropy density. On the contrary, hexagonality seems to be a function of excess entropy. A maximum value of hexagonality is found for an excess entropy of $0.8724$ bits.

%-------------------------------------------------------------------------------------------------------------------
\subsection{The pairwise correlation function.}
%-------------------------------------------------------------------------------------------------------------------

The HMM for the $ABC$ coding describing the extrinsic fault can be constructed from the H\"agg description and is shown in Fig. \ref{fig:abcfsa}. For each state in the HMM over the H\"agg code (Fig. \ref{fig:fsaext}), three states are induced in the HMM of the $ABC$ coding corresponding to subsequences starting with $A$, $B$ and $C$. Using the same procedure described for the H\"agg HMM, the transition matrices can be written and the stationary probability over the recurrent states calculated  for the HMM over the $ABC$ coding:
\begin{equation}
 \langle \pi_{abc}|=\frac{1}{3(1+\gamma)}\{\gamma, \gamma, \gamma,1,1,1\}.\label{eq:piabc}
\end{equation}
where the order in the states has been taken as $\{A_e, B_e, C_e, A, B, C\}$. Using equation (\ref{eq:Q}) the pairwise correlation follows
\begin{equation}
\begin{array}{l}
\displaystyle Q_{s}(\Delta)=\frac{1}{3}\left [ 1+\right.\\\\
\left  (\frac{|p|}{4}\right )^{\Delta}\left(\left[1+\frac{\cos(3 \phi_r)|r|}{\sqrt{3}(1+\gamma)}\right]\cos(\Delta\phi_p)+\frac{\sin(3\phi_r)|r|}{\sqrt{3}(1+\gamma)}\sin(\Delta \phi_p)\right)+\\\\
\left.\left  (\frac{|q|}{4}\right )^{\Delta}\left(\left[1-\frac{\cos(3 \phi_r)|r|}{\sqrt{3}(1+\gamma)}\right]\cos(\Delta\phi_q)-\frac{\sin(3\phi_r)|r|}{\sqrt{3}(1+\gamma)}\sin(\Delta \phi_q)\right)\right ]\\\\
=\frac{1}{3}\left( 1+Q^{[1]}_{s}(\Delta)+Q^{[2]}_{s}(\Delta)\right ), \label{eq:q0}
\end{array}
\end{equation}
where
\[\begin{array}{l}
r=|r|e^{i \phi_r}=\sqrt{i \sqrt{3}(6\gamma-\gamma^2-1)-(1+\gamma)^2},\\\\
x=(\gamma-1)(1-i\sqrt{3}),\\\\
p=|p|e^{\phi_p}=x+\sqrt{2}s,\\\\
q=|q|e^{\phi_q}=x-\sqrt{2}s.
\end{array}
\] 

The obtained equation is equivalent to that result given by \cite{holloway69}, as can be seen by comparing numerical results from equation (\ref{eq:q0}) for $\Delta=0,1,2,3$ and those reported in equations (35), (36), (37), (38) in  \cite{holloway69a} (making $\alpha=0$)\footnote{When comparing with \cite{holloway69a} results, it must be noticed that in their notation $Q_{s}(\Delta)=P(m)$, $Q_{c}(\Delta)=Q(m)$ and $Q_{a}(\Delta)=R(m)$}. In turn these authors have shown that their result reduces to that of \cite{johnson63}. Holloway and Klamkin do not give a close form  of $Q_{s}(\Delta)$ for $\Delta > 3$. 

There are two terms in the expression for $Q_{s}(\Delta)$, each with an oscillating and a decaying part. Figure \ref{fig:qp}a shows the behavior of both decaying terms with faulting probability. $p$ and $q$ have a jump (discontinuity) at the same value of $\gamma \approx \gamma_0= 0.1716$, where the real part of $r$ has a minimum, and the imaginary part has a jump from negative value to a positive one. Interesting, the combined plot of both terms result in two smooth continuous curves. At $\gamma=0$, $p$ is zero while $q=1$, the oscillating part of the second term in $Q_{s}$ dominates. At $\gamma=1$, both $p$ and $q$ have the same value of $1$ and the combine effect of both oscillating terms determines the pairwise correlation function. for both cases( $\gamma=0$ and  $\gamma=1$), $Q_{s}(\Delta)$ reduces to 
\[
\displaystyle Q_{s}(\Delta)=\frac{1}{3}\left(1+2 \cos \left[\frac{2 \pi}{3}\Delta\right] \right ),
\]
describing the correlation function for the perfect $3C$ stacking.

At $\gamma_0$ the oscillating part of both terms in $Q_{s}(\Delta)$ becomes equal for all values of $\Delta$. At $\gamma=\sqrt{2}-1$, where the hexagonality reaches its maximum value, the oscillating part of $Q^{[1]}_{s}(\Delta)$ is the prevailing one at large $\Delta$ values. For small ($\gamma\approx 0$) and large values ($\gamma\approx 1$) is the oscillating part of $Q^{[2]}_{s}(\Delta)$ which determines the underlying stacking sequence.

In any case, the lower curve in Fig. \ref{fig:qp}a determines the faster decaying behavior of the pairwise correlation function, while the upper curve determines the dominant behavior at larger $\Delta$ values. Figure \ref{fig:qp}b shows the correlation lengths derived from both decaying terms.  At large values of $\Delta$ the $p$ term is the dominant factor in the pairwise correlation function for values of $\gamma > \gamma_0$, while the opposite happens at values below $\gamma_0$. 

A similar deduction made for $Q_{c}(\Delta)$ results in 
\[\begin{array}{l}
Q_c(\Delta)=\frac{1}{3} \left(1+\left ( \frac{\left| p\right|}{4}\right )^\Delta \left [C_p \cos (\Delta \phi_p)+S_p \sin (\Delta \phi_p)\right]+\right.\\\\
\left.\left ( \frac{\left| q\right|}{4}\right )^\Delta  \left[C_q \cos (\Delta\phi_q)+S_q \sin (\Delta \phi_q)\right]\right)
,\end{array}\]
with
\[
\begin{array}{l}
 C_p=\frac{\sqrt{2}}{\left| r \right|}\frac{ \gamma^2-4 \gamma+1 }{1+\gamma  }\cos\phi_r+2 \frac{\sqrt{6}}{\left| r \right|}\frac{\gamma}{1+\gamma}\sin \phi_r-\frac{1}{2},\\\\
 S_p=\frac{\sqrt{2}}{\left| r \right|}\frac{ \gamma^2-4 \gamma+1 }{1+\gamma}\sin\phi_r-2 \frac{ \sqrt{6}}{\left| r \right|}\frac{\gamma}{1+\gamma} \cos \phi_r+\frac{\sqrt{3}}{2}\\\\
 C_q=-\frac{\sqrt{2}}{\left| r \right|}\frac{ \gamma^2-4 \gamma+1 }{1+\gamma  }\cos\phi_r-2 \frac{\sqrt{6}}{\left| r \right|}\frac{\gamma}{1+\gamma}\sin \phi_r-\frac{1}{2},\\\\
 S_q=-\frac{\sqrt{2}}{\left| r \right|}\frac{ \gamma^2-4 \gamma+1 }{1+\gamma}\sin\phi_r+2 \frac{ \sqrt{6}}{\left| r \right|}\frac{\gamma}{1+\gamma} \cos \phi_r+\frac{\sqrt{3}}{2}.
 \end{array}
\]

$Q_a(\Delta)$ follows from the normalization condition.

%-------------------------------------------------------------------------------------------------------------------
\section{The interference function.}
%-------------------------------------------------------------------------------------------------------------------

The diffraction pattern of an OD structure can be decomposed in two contributions: that of the layer and that of the stacking sequence. The reduced diffracted intensities (i.e. once the necessary corrections are applied: Lorentz, polarization, absorption etc.) can be deconvoluted in terms of these two contributions so that the stacking sequence leaves its fingerprint in the form of an interference function showing a periodic distribution of deconvoluted intensities.

In the case of complex sequences like that of micas, in which adjacent layers can be stacked in six different orientations, the interference function has been called PID (Periodic Intensity Distribution: \cite{nespolo99}). For close packed structures, the situation is simpler because adjacent layers may take only two relative positions. The consequence of extrinsic faulting over the diffracted intensity is visible in the interference function. The interference function  follows from the use of the expressions for $Q_{s}$, $Q_{c}$ and $Q_{a}$ \cite{estevez01}:

\begin{equation}
 \displaystyle {\cal I}({r}^{*})= 1+2  \sum_{\Delta=1}^{N_{c}-1} A_{\Delta} \cos(2 \pi \Delta l)+B_{\Delta} \sin(2 \pi \Delta l), \label{Qfinal} 
 \end{equation}
where
\begin{equation}
\label{fcoef}
\begin{array}{l}
\displaystyle A_{\Delta}=(1-\frac{\Delta}{N_{c}}) \left \{ Q_s(\Delta) +\left[Q_c(\Delta)+Q_a(\Delta)\right] \cos[\frac{2 \pi}{3} (h-k)]\right \}\label{fcoefa}\\\\
\displaystyle B_{\Delta}=(1-\frac{\Delta}{N_{c}}) \left[Q_c(\Delta)-Q_a(\Delta)\right] \sin[\frac{2 \pi}{3} (h-k)].
\end{array}
\end{equation}
$N_c$ is the number of layers in the stacking sequence.

For $h-k$ a multiple of $3$, the coefficients reduces to $A_{\Delta}=(1-\frac{\Delta}{N_{c}})$ and $B_{\Delta}=0$ and this family of reflections are not affected by the extrinsic faulting. For $h-k=3n+1$ with $n$ an integer, the coefficients are then 
\begin{equation}\label{fcoef1}
\begin{array}{l}
\displaystyle A_{\Delta}=(1-\frac{\Delta}{N_{c}}) \left \{ Q_s(\Delta) -\frac{Q_c(\Delta)+Q_a(\Delta)}{2}\right \}\\\\
\displaystyle B_{\Delta}= \frac{\sqrt{3}}{2}(1-\frac{\Delta}{N_{c}})\left[Q_c(\Delta)-Q_a(\Delta)\right] .
\end{array}
\end{equation}
The last case is $h-k=3n+2$ with $n$ an integer, the coefficients are then 
\begin{equation}\label{fcoef2}
\begin{array}{l}
\displaystyle A_{\Delta}=(1-\frac{\Delta}{N_{c}}) \left \{ Q_s(\Delta) -\frac{Q_c(\Delta)+Q_a(\Delta)}{2}\right \}\\\\
\displaystyle B_{\Delta}= \frac{\sqrt{3}}{2}(1-\frac{\Delta}{N_{c}})\left[Q_a(\Delta)-Q_c(\Delta)\right] .
\end{array}
\end{equation}

An analytical expression for the interference function can be deduced from the above equations but is too long and cumbersome to be of any particular interest\footnote{In  Riechers, Varn, and Crutchfield arXiv:1410.5028 (2014), a more elegant way to deduce the interference function directly from the HMM is derived and could lead to a more manageable expression as has been rightly pointed out by an anonymous referee.}.

The result has been discussed already by \cite{johnson63,warren63,holloway69a}. With increasing faulting probability $\gamma$, the peak asymmetrically broadens, lowers its intensity and shifts (Figure \ref{fig:peakshift}). For $h-k=1\;(mod \, 3)$ the peak originally at $l=3n+1$ ($n\in \mathcal{Z}$) shift towards lower $l$ values, while the opposite occurs for $h-k=2\;(mod \, 3)$ where the peak originally at $l=3n-1$  shift towards higher $l$ values. Additionally at high faulting probability an additional peak appears near the so called twin position. For  $h-k=1\;(mod \, 3)$ ($h-k=2\;mod \, 3$) the twin position is at $l=3n-1$ ($l=3n+1$), the additional peak appears at lower (larger) value of $l$ and gradually shifts towards the twin peak position as $\gamma$ increases while strengthening its intensity and decreasing its broadening. The behavior of the original peak and the twin one are not symmetrical, that is, they do not behave the same for $\gamma$ and $1-\gamma$, respectively. The non symmetric behavior of the peaks can be explained by the non symmetrical character of the HMM describing extrinsic broadening (Figure \ref{fig:fsaext}). A similar profile for single crystal and the particular case of $\gamma=1/2$ has been reported by \cite{varn13}.

If one observe the interference for $\gamma=0.333$ (Figure \ref{fig:peakshift}), it is too often in the literature that peak deformations with geometry such as these are fitted with models involving more than one phase. The fact that such distortions can be result of a single type of faulting that does not lead to any polytype, should be taken as a note of warning against introducing to easily new structures in profile fitting.

In Figure \ref{fig:asym} the peak shift and asymmetry as a function of faulting probability is shown. Asymmetry has been defined as the ratio between the half width at half maximum (HWHM) for the right side ($W_r$), divided  by the HWHM for the left side ($W_l$), by construction, the asymmetry is equal to $1$ for a perfect symmetric peak.

For powder diffraction it must be considered that the components of a family of planes like the $\{111\}$ ( where all members of the family are crystallographically  equivalent for the unfaulted crystal and share the same interplanar distance) are no longer equivalent when faulting occurs. For example, when indexed respect to hexagonal axis the $\{111\}$ includes the following planes: $(0,0,3)$ , $(0,0,\bar{3})$, $(\bar{1},1,1)$, $(1,0,1)$, $(0,\bar{1},1)$, $(0,1,\bar{1})$, $(1,\bar{1},\bar{1})$, $(\bar{1},0,\bar{1})$; the first two are unaffected by extrinsic faulting, the next three are of the type $h-k=1\;(mod\, 3)$, and the last three of the form $h-k=2\;(mod\, 3)$. Thus, when simulating the faulted powder diffraction profiles, each component of a plane family must be considered individually. Figure \ref{fig:powder} shows the powder peak profile for the  $\{111\}$ where the components not affected by faulting have been left out. The reader can compare with the single crystal profiles of figure \ref{fig:peakshift}.  

%-------------------------------------------------------------------------------------------------------------------
\section{Conclusions}
%-------------------------------------------------------------------------------------------------------------------

Stacking disorder can be viewed in a number of cases as a dynamical system capable of storing and processing information. From this point of view, it has been shown that extrinsic fault in the H\"agg code is a sofic system, where predictability in the future is linked to long range memory in the past for faulting probabilities within ]0,1[. A sofic system, as the one considered here, has no description as a finite range Markov process. This inability to describe such simple faulting process by a finite range model is interesting, as it is common in the literature to try to model faulting by this type of finite range Markov models.\footnote{We thank one of the anonymous referee for her/his enlightening comment on this issue}.  In spite of this, the HMM model for extrinsic faulting is simple enough, it just belongs to different type of processing machinery. This is precisely the underlying idea of computational mechanics that attempts to find the less sophisticated model for a given process by climbing up in a given hierarchy of possible computational machines until such description is found.

This character has several interesting consequences. First, the excess entropy equals the statistical complexity of the system. Excess entropy is linked with the structured output of the system, while statistical complexity measures memory stored in the system. In consequence, structure is linked to memory, a result not surprising once it is acknowledge that the HMM of the process is equivalent to a biased even process. In an even process, the occurrence of consecutive 0's has to be tracked completely, to determine in which state the system is. As increasing faulting probability means longer runs of 0's, excess entropy grows monotonically with increasing $\gamma$. Excess entropy has a discontinuity at $\gamma=1$, where the topology of the HMM changes to a one state system with certainty in the output and therefore zero $E$. 

Entropy density, on the other hand, is a smooth function of faulting probability in all the probability range. Entropy density has a maximum at $\gamma \approx 0.382$, near the maximum of the hexagonality, but slightly larger value. Extrinsic faulting, as treated here, implies that no faulting probability changes the underlying periodic sequence: no phase transformation happens. Hexagonality reaches at $\gamma =\sqrt{2}-1\approx 0.414214$ a maximum of $2(3-2\sqrt{2})\approx0.34314$  and therefore the system is always more ``cubic'' than hexagonal. 

In the text, several useful analytical expressions have been derived for different entropic magnitudes, probabilities, lengths, and correlations, all as a function of the faulting probability $\gamma$. To the knowledge of the authors, such expressions have not been reported before. 

The pairwise correlation function of the layers has been derived and from there the interference function was obtained. The correlation function is composed of two terms each with a decaying and oscillating part. The numerical values of the obtained expression coincides with those that can be found using previous treatments. The shift and asymmetric broadening of the reflections as a result of extrinsic faulting was also discussed.

\section{Acknowledgment}

This work was partially financed by FAPEMIG under the project BPV-00047-13 and computational infrastructure support under project APQ-02256-12. EER which to thank the Universit\'e de Lorraine for a visiting professor grant. He also would like to acknowledge the financial support under the PVE/CAPES grant 1149-14-8 that allowed the visit to the UFU.  RLS wants to thank the support of CNPq through the projects 309647/2012-6 and 304649/2013-9. We would like to thank the anonymous referees for the careful reading and the number of valuable suggestion that greatly improved the manuscript.

%*************************************************************************************
\section{Appendix}
%***********************************************************************

\subsection{Calculation of the excess entropy}

In order to calculate the excess entropy the mixed state representation of the system dynamics must be deduced. To understand what the mixed state representation is, the HMM description must be viewed as an instance of a  hidden Markov model (HMM) \cite{upper89}. In short, any model derived by the observer of the system output, that reproduces (statistically) the output, is called a presentation of the process. The observer can then follow the evolution of the system by updating mixed states, defined as a distribution over the states of the HMM HMM description. The reader is referred to \cite{upper89} and \cite{crutchfield13} for a detailed explanation, the later will closely followed. 

\begin{figure}
\caption{Finite state automata (HMM) over the H\"agg code for the extrinsic fault in the $3C$ stacking. For $\gamma=0$ and $\gamma=1$ the HMM describes a perfect $3C$ stacking. The output of the HMM will be of the perfect $3C$ structure as long as it stays in the $p$ state, a transition to the $e$ state signals the occurrence of an extrinsic fault described by, at least, a pair of $0's$. Observe that the perfect sequence is described by the cyclic $ABC$ sequence corresponding to $1's$ in the H\"agg notation, an equivalent HMM could be described for the anti-cyclic $CBA$ sequence by a trivial relabeling of the output symbols.}
\scalebox{1.5}{\includegraphics{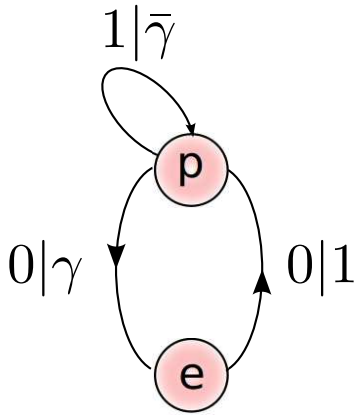}}
\label{fig:fsaext}
\end{figure}

\begin{figure}
\caption{(a) Hexagonality, entropy density  and (b) excess entropy as a function of the faulting probability. Excess entropy has a discontinuity at $\gamma=1$, where it drops to zero.}
\scalebox{1.5}{\includegraphics{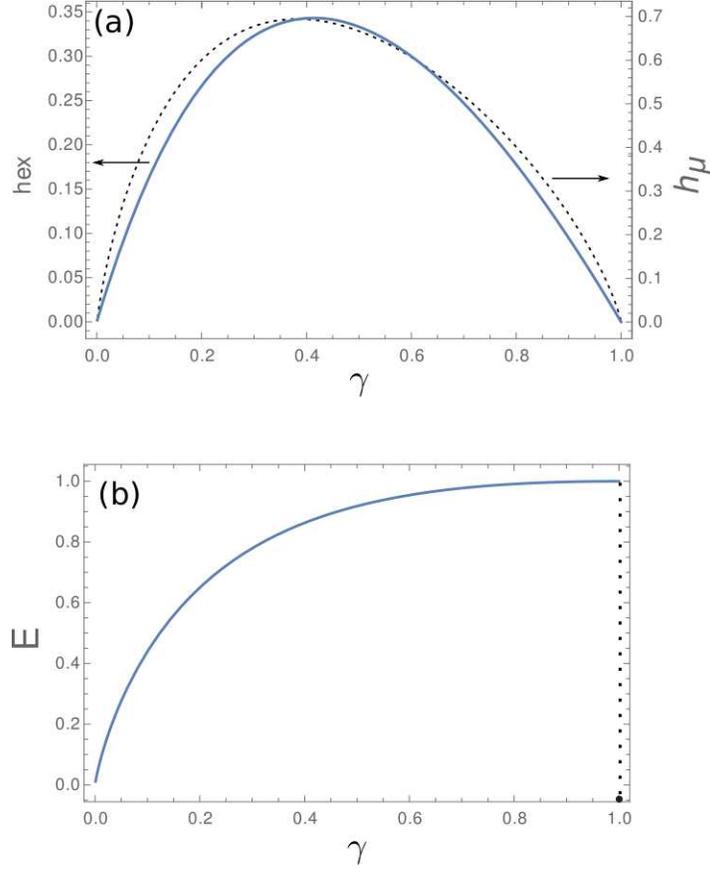}}
\label{fig:hexhe}
\end{figure}

\begin{figure}
\caption{Hexagonality as a function of (a) entropy density, (b) excess entropy.}
\scalebox{0.7}{\includegraphics{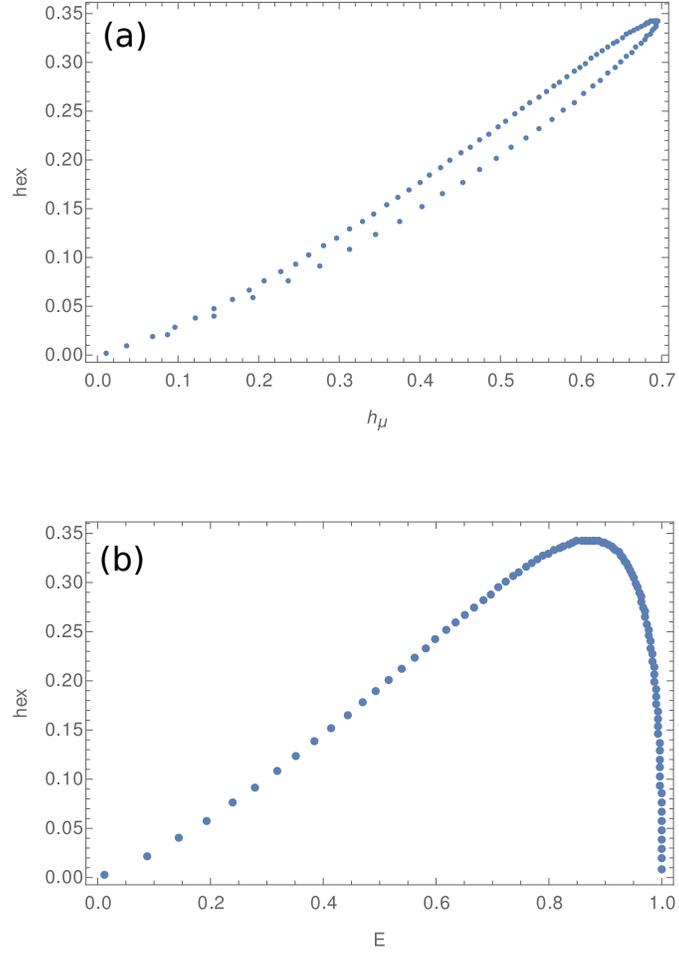}}
\label{fig:hexvse}
\end{figure}

\begin{figure}
\caption{Finite state automata over the ABC code for the extrinsic fault in the $3C$ stacking. The HMM can be derived from that of Fig. \ref{fig:fsaext}. The subscript in the states of the HMM are induced by the states with the same label in the H\"agg HMM.}
\scalebox{1}{\includegraphics{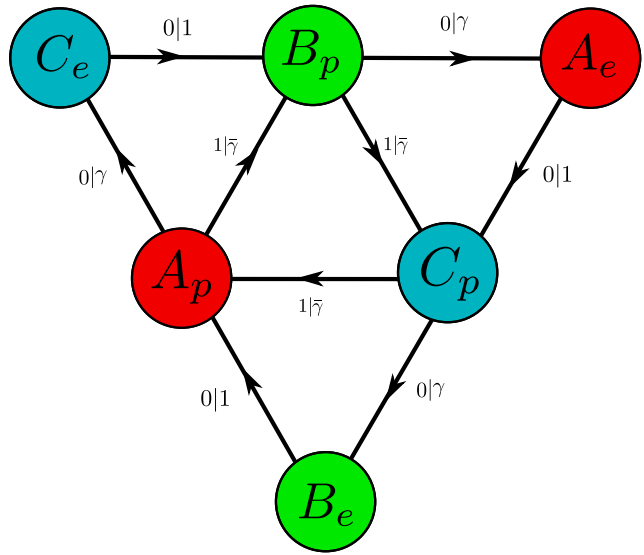}}
\label{fig:abcfsa}
\end{figure}

\begin{figure}
\caption{(a) $Q_{s}$ decaying terms $p/4$ and $q/4$, taken from expression (\ref{eq:Q}) as a function of faulting probability. The decaying terms are discontinuous at $\gamma=0.1716$ where, with increasing value of $\gamma$, $p/4$ has a jump to larger value, while $q/4$ jumps to a lower value. Yet, the combined plot of both functions results in two smooth continuous curves. (b) The correlation length for both decaying terms.}
\scalebox{2.0}{\includegraphics{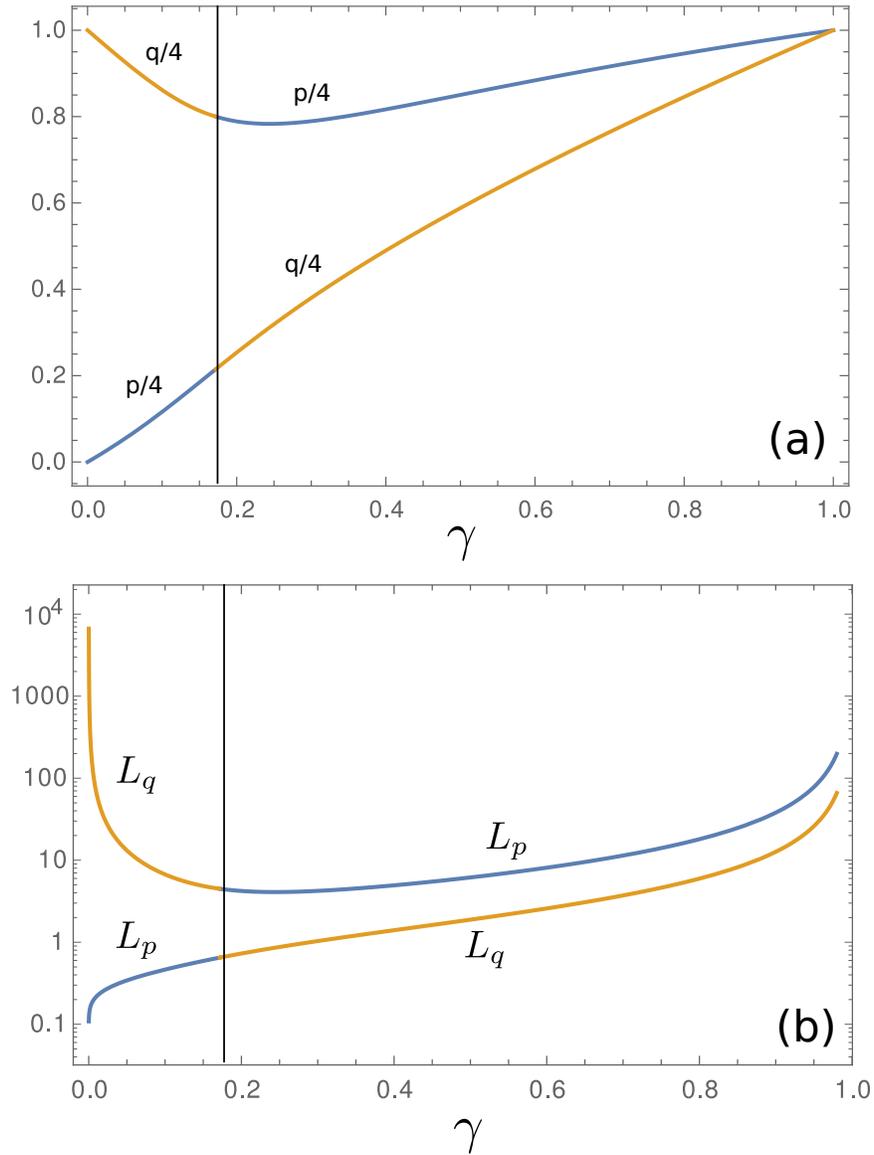}}
\label{fig:qp}
\end{figure}

\begin{figure}
\caption{The mixed state representation of the biased even process. The starting state is denoted by the double circle. Compare with figure 2 of the appendix A in Crutchfield et al. (2013).}
\scalebox{1.5}{\includegraphics{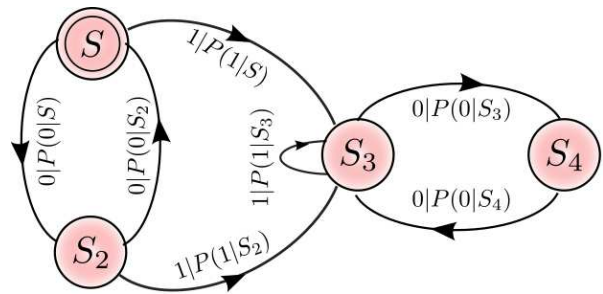}}
\label{fig:mixedfsm}
\end{figure}

\begin{figure}
\caption{The behavior of the peaks for different extrinsic faulting probability. Vertical dashed line marks the unfaulted peak positions. For $h-k=2\;(mod\, 3)$ the pattern is similar just reflected over the $l=1.5$ vertical line. Simulation was performed directly from the interference function.}
\scalebox{1.5}{\includegraphics{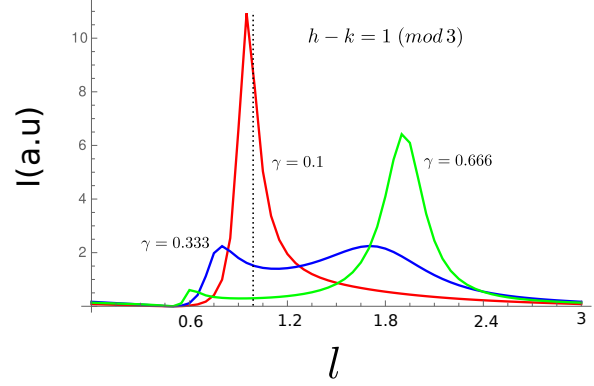}}
\label{fig:peakshift}
\end{figure}

\begin{figure}
\caption{(a) Peak shift as a function of faulting probability. $l_m$ is the position of the peak maximum. Dashed lines mark the unfaulted peak position. (b) Asymmetry of the peaks with $h-k=1\;(mod \,3)$ as a function of faulting probability for small values of $\gamma$. Asymmetry is defined as the ratio between the HWHM at the right side divided by the HWHM at the left side.}
\scalebox{2.5}{\includegraphics{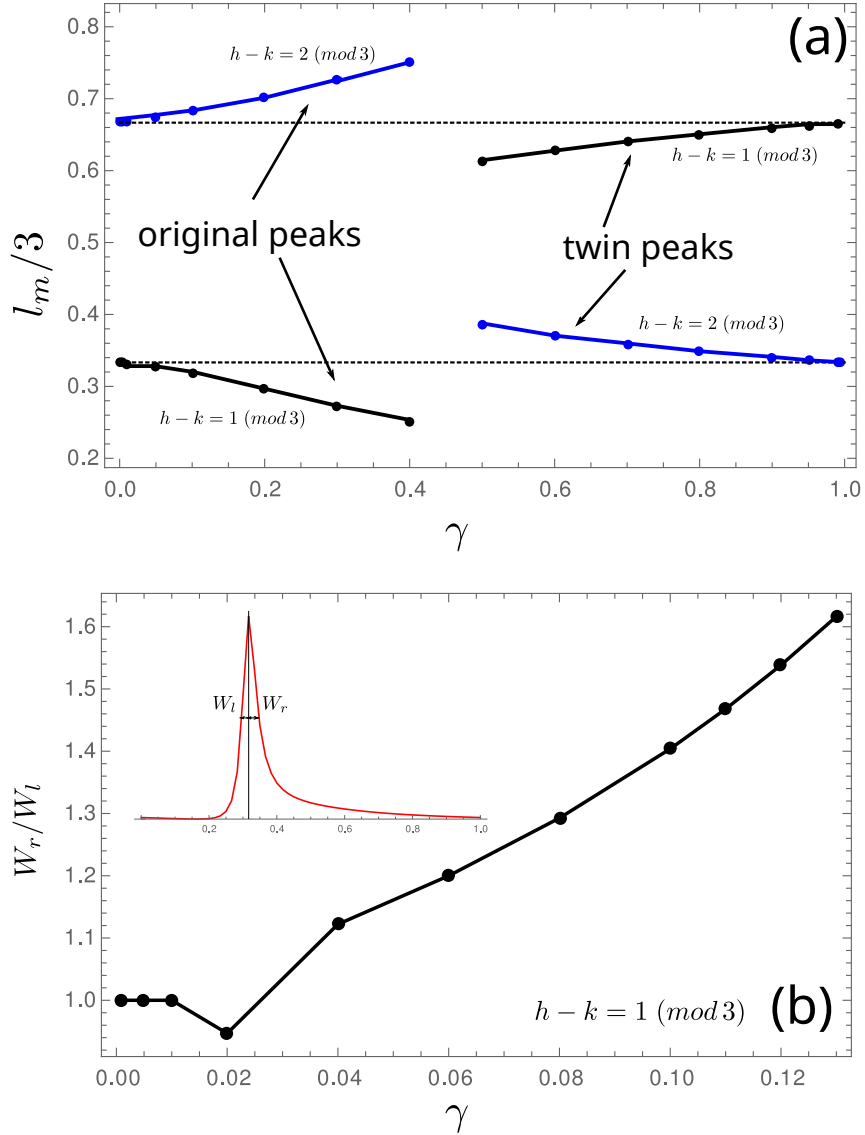}}
\label{fig:asym}
\end{figure}

\begin{figure}
\caption{The behavior of the $\{111\}$ peaks for a powder diffraction pattern for different extrinsic faulting probability. The peaks not affected by faulting have been left out.}
\scalebox{1.0}{\includegraphics{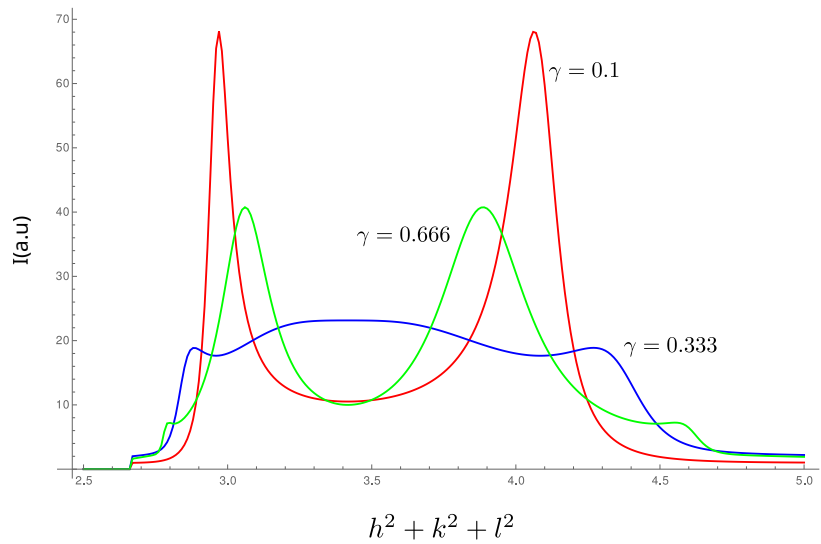}}
\label{fig:powder}
\end{figure}

The mixed state representation of the biased even process of Figure \ref{fig:fsaext} is shown in Figure \ref{fig:mixedfsm} (Compare with Fig. 2 in \cite{crutchfield13}). Each state in the set $\mathcal{S}$ now has a distribution of probabilities associated with it:
\[
 \begin{array}{ll}
  S: & \delta_S= \left \{\frac{1}{1+\gamma}, \frac{\gamma}{1+\gamma}   \right \}\\\\
  S_2: & \delta_{S_2}=\left \{1/2, 1/2   \right \}\\\\
  S_3: & \delta_{S_3}=\left \{1,0 \right \}\\\\
  S_4: & \delta_{S_4}=\left \{0, 1\right \}.
 \end{array}
\]
As well as transition probabilities
\[
 \begin{array}{ll}
  P(0|S)= & 2 \frac{\gamma}{1+\gamma}\\\\
  P(1|S)= & \frac{1-\gamma}{1+\gamma}\\\\
  P(0|S_2)= & \frac{1+\gamma}{2}\\\\
  P(1|S_2)= & \frac{1-\gamma}{2}\\\\
  P(0|S_3)= & 1-\gamma\\\\
  P(1|S_3)= & \gamma\\\\
  P(0|S_4)= & 1\\\\
  P(1|S_4)= & 0\\\\
 \end{array}
\]

Observe that the emission of a $1$ implies from any state, a transition to the state $S_3$. States $S$ and $S_2$ are transient while the recurrent states reproduce the original HMM. The stationary probability over the states is given by
\[
 \displaystyle \langle \pi_{mix}|=\left \{0,0,\frac{1}{1+\gamma}, \frac{\gamma}{1+\gamma}\right \}.
\]
The state transition matrix will be
\[
\displaystyle W=\left (
 \begin{array}{cccc}
  0                  & \frac{2\gamma}{1+\gamma}& \frac{1-\gamma}{1+\gamma} & 0\\
  \frac{1+\gamma}{2} & 0                       & \frac{1-\gamma}{2}        & 0 \\
  0                  & 0                       & 1-\gamma                  & \gamma \\
  0                  & 0                       & 1                         & 0
 \end{array}
 \right ).
\]
With eigenvalues
\[
 \Lambda_{W}=\left \{ 1,-\gamma, -\sqrt{\gamma}, \sqrt{\gamma} \right \}.
\]
The projection operator $W_{\lambda}$, for each eigenvalue, is obtained using
\[
 \displaystyle W_\lambda=\prod_{\xi\in\Lambda_W,\xi \neq \lambda}\frac{W-\xi I}{\lambda-\xi}
\]
$I$ represents the identity matrix and the product avoids the singularity in the denominator. The results are

\[
\begin{array}{l}
\displaystyle W_1=\left (
 \begin{array}{cccc}
  0  & 0 & \frac{1}{1+\gamma}& \frac{\gamma}{1+\gamma}\\
  0  & 0 & \frac{1}{1+\gamma}& \frac{\gamma}{1+\gamma}\\
  0  & 0 & \frac{1}{1+\gamma}& \frac{\gamma}{1+\gamma}\\
  0  & 0 & \frac{1}{1+\gamma}& \frac{\gamma}{1+\gamma}
 \end{array}
\right )\\\\
\displaystyle W_{-\gamma}=\left (
 \begin{array}{cccc}
  0  & 0 & 0                 & 0\\
  0  & 0 & \frac{1}{2}\frac{\gamma-1}{1+\gamma}& \frac{1}{2}\frac{1-\gamma}{1+\gamma}\\
  0  & 0 & \frac{\gamma}{1+\gamma}& -\frac{\gamma}{1+\gamma}\\
  0  & 0 &- \frac{1}{1+\gamma}& \frac{1}{1+\gamma}
 \end{array}
\right )\\\\
\displaystyle W_{-\sqrt{\gamma}}=\left (
 \begin{array}{cccc}
  \frac{1}{2}  & -\frac{\sqrt{\gamma}}{1+\gamma} & \frac{1}{2}\frac{\sqrt{\gamma}-1}{1+\gamma}& \frac{1}{2}\frac{\sqrt{\gamma}-\gamma}{1+\gamma}\\
  -\frac{1}{4}\frac{1+\gamma}{\sqrt{\gamma}}  & \frac{1}{2} & \frac{1}{4}\frac{1-\sqrt{\gamma}}{\sqrt{\gamma}}& \frac{1}{4}(\sqrt{\gamma}-1)\\
  0  & 0 & 0& 0\\
  0  & 0 & 0&0
 \end{array}
\right )\\\\
\displaystyle W_{\sqrt{\gamma}}=\left (
 \begin{array}{cccc}
  \frac{1}{2}  & \frac{\sqrt{\gamma}}{1+\gamma} & -\frac{1}{2}\frac{\sqrt{\gamma}+1}{1+\gamma}& -\frac{1}{2}\frac{\sqrt{\gamma}+\gamma}{1+\gamma}\\
  \frac{1}{4}\frac{1+\gamma}{\sqrt{\gamma}}  & \frac{1}{2} & -\frac{1}{4}\frac{1+\sqrt{\gamma}}{\sqrt{\gamma}}& -\frac{1}{4}(\sqrt{\gamma}+1)\\
  0  & 0 & 0& 0\\
  0  & 0 & 0&0
 \end{array}
\right ).
\end{array}
\]
Defining
\[
\langle \delta_\pi|=\{ \begin{array}{llll}1 & 0 & 0 & 0\end{array}\}
\]
then
\[
| H(W^{\mathcal{A}})\rangle=-\sum_{\eta \in \mathcal{S}}|\delta_{\eta}\rangle\sum_{x\in \{0,1\}}\langle \delta_\eta|W^{(x)} |\mathbf{1}\rangle \log \langle \delta_\eta|W^{(x)} |\mathbf{1}\rangle,
\]
and the excess entropy follows from
\[
E=\sum_{\lambda\in\Lambda_W, |\lambda|<1}\frac{1}{1-\lambda}\langle \delta_{\pi_{mix}}|W_\lambda|H(W^{\mathcal{A}})\rangle
\]
which is equation (8) from \cite{crutchfield13}.

%\referencelist[extrinsic]

%***********************************************************************
\newpage

%\end{multicols}

\end{document}